\documentclass[12pt]{article}
\begin{document}
\newcommand{\beq}{\begin{equation}}
\newcommand{\eeq}{\end{equation}}
\newcommand{\beqa}{\begin{eqnarray}}
\newcommand{\eeqa}{\end{eqnarray}}
\newcommand{\beqar}{\begin{eqnarray*}}
\newcommand{\eeqar}{\end{eqnarray*}}
\newcommand{\al}{\alpha}
\newcommand{\be}{\beta}
\newcommand{\del}{\delta}
\newcommand{\D}{\Delta}
\newcommand{\eps}{\epsilon}
\newcommand{\ga}{\gamma}
\newcommand{\Ga}{\Gamma}
\newcommand{\ka}{\kappa}
\newcommand{\nn}{\nonumber}
\newcommand{\inn}{\!\cdot\!}
\newcommand{\h}{\eta}
\newcommand{\ii}{\iota}
\newcommand{\kk}{\varphi}
\newcommand\F{{}_3F_2}
\newcommand{\la}{\lambda}
\newcommand{\La}{\Lambda}
\newcommand{\na}{\prt}
\newcommand{\Om}{\Omega}
\newcommand{\om}{\omega}
\newcommand{\p}{\phi}
\newcommand{\sig}{\sigma}
\renewcommand{\t}{\theta}
\newcommand{\z}{\zeta}
\newcommand{\ssc}{\scriptscriptstyle}
\newcommand{\eg}{{\it e.g.,}\ }
\newcommand{\ie}{{\it i.e.,}\ }
\newcommand{\labell}[1]{\label{#1}} 
\newcommand{\reef}[1]{(\ref{#1})}
\newcommand\prt{\partial}
\newcommand\veps{\varepsilon}
\newcommand{\pol}{\varepsilon}
\newcommand\vp{\varphi}
\newcommand\ls{\ell_s}
\newcommand\cF{{\cal F}}
\newcommand\cA{{\cal A}}
\newcommand\cS{{\cal S}}
\newcommand\cT{{\cal T}}
\newcommand\cV{{\cal V}}
\newcommand\cL{{\cal L}}
\newcommand\cM{{\cal M}}
\newcommand\cN{{\cal N}}
\newcommand\cG{{\cal G}}
\newcommand\cH{{\cal H}}
\newcommand\cI{{\cal I}}
\newcommand\cJ{{\cal J}}
\newcommand\cl{{\iota}}
\newcommand\cP{{\cal P}}
\newcommand\cQ{{\cal Q}}
\newcommand\cg{{\it g}}
\newcommand\cR{{\cal R}}
\newcommand\cB{{\cal B}}
\newcommand\cO{{\cal O}}
\newcommand\tcO{{\tilde {{\cal O}}}}
\newcommand\bg{\bar{g}}
\newcommand\bb{\bar{b}}
\newcommand\bH{\bar{H}}
\newcommand\bX{\bar{X}}
\newcommand\bK{\bar{K}}
\newcommand\bR{\bar{R}}
\newcommand\bZ{\bar{Z}}
\newcommand\bxi{\bar{\xi}}
\newcommand\bphi{\bar{\phi}}
\newcommand\bpsi{\bar{\psi}}
\newcommand\bprt{\bar{\prt}}
\newcommand\bet{\bar{\eta}}
\newcommand\btau{\bar{\tau}}
\newcommand\hF{\hat{F}}
\newcommand\hA{\hat{A}}
\newcommand\hT{\hat{T}}
\newcommand\htau{\hat{\tau}}
\newcommand\hD{\hat{D}}
\newcommand\hf{\hat{f}}
\newcommand\hg{\hat{g}}
\newcommand\hp{\hat{\phi}}
\newcommand\hi{\hat{i}}
\newcommand\ha{\hat{a}}
\newcommand\hb{\hat{b}}
\newcommand\hQ{\hat{Q}}
\newcommand\hP{\hat{\Phi}}
\newcommand\hS{\hat{S}}
\newcommand\hX{\hat{X}}
\newcommand\tL{\tilde{\cal L}}
\newcommand\hL{\hat{\cal L}}
\newcommand\tG{{\widetilde G}}
\newcommand\tg{{\widetilde g}}
\newcommand\tphi{{\widetilde \phi}}
\newcommand\tPhi{{\widetilde \Phi}}
\newcommand\te{{\tilde e}}
\newcommand\tk{{\tilde k}}
\newcommand\tf{{\tilde f}}
\newcommand\ta{{\tilde a}}
\newcommand\tb{{\tilde b}}
\newcommand\tR{{\tilde R}}
\newcommand\teta{{\tilde \eta}}
\newcommand\tF{{\widetilde F}}
\newcommand\tK{{\widetilde K}}
\newcommand\tE{{\widetilde E}}
\newcommand\tpsi{{\tilde \psi}}
\newcommand\tX{{\widetilde X}}
\newcommand\tD{{\widetilde D}}
\newcommand\tO{{\widetilde O}}
\newcommand\tS{{\tilde S}}
\newcommand\tB{{\widetilde B}}
\newcommand\tA{{\widetilde A}}
\newcommand\tT{{\widetilde T}}
\newcommand\tC{{\widetilde C}}
\newcommand\tV{{\widetilde V}}
\newcommand\thF{{\widetilde {\hat {F}}}}
\newcommand\Tr{{\rm Tr}}
\newcommand\tr{{\rm tr}}
\newcommand\STr{{\rm STr}}
\newcommand\hR{\hat{R}}
\newcommand\M[2]{M^{#1}{}_{#2}}

\newcommand\bS{\textbf{ S}}
\newcommand\bI{\textbf{ I}}
\newcommand\bJ{\textbf{ J}}

\begin{titlepage}
\begin{center}

\vskip 2 cm
{\LARGE \bf    Effective action of bosonic string theory   \\  \vskip 0.25 cm  at order $\alpha'^2$
 }\\
\vskip 1.25 cm
   Mohammad R. Garousi\footnote{garousi@um.ac.ir}

\vskip 1 cm
{{\it Department of Physics, Faculty of Science, Ferdowsi University of Mashhad\\}{\it P.O. Box 1436, Mashhad, Iran}\\}
\vskip .1 cm
 \end{center}

\begin{abstract}
 Recently, it has been shown  that  the  gauge  invariance requires the minimum number of  independent couplings  for   $B$-field, metric and dilaton at order $\alpha'^2$ to be 60. In this paper we fix the corresponding  60 parameters in string theory by requiring the couplings to be invariant under the global T-duality transformations. The Riemann cubed terms  are exactly the same as the couplings that  have been found by the S-matrix calculations.
  
\end{abstract}
\end{titlepage}

\section{Introduction}
String theory is a quantum theory of gravity   with  a finite number of massless fields and a  tower of infinite number of  massive fields reflecting the stringy nature of the gravity.
An efficient way to study different phenomena in this theory is to use an effective action  which includes     only  massless fields. The effects of the massive fields appear in the action   as the  higher derivatives of the massless fields. This effective action may be found  by imposing various  symmetries/dualities in the string theory. There are various gauge symmetries in the effective actions which  are corresponding to the various massless fields, \eg the diffeomorphism symmetry corresponds to the metric and  the  gauge symmetry corresponds to the Kalb-Ramond field or $B$-field. In the bosonic string theory which has only metric, dilaton and $B$-field, they are the only local symmetries of the effective action. Imposing only these symmetries, one finds the effective action   has three couplings at order $\alpha'^0$ (two-derivative order),   has 8 couplings at order $\alpha'$ (four-derivative order)   up to field redefinitions \cite{Metsaev:1987zx},  has 60 couplings at order $\alpha'^2$ (six-derivative order) \cite{Garousi:2019cdn}  and so on. The gauge symmetries, however, can not determine the coefficients of the couplings. These parameters  may be found by S-matrix calculations \cite{Scherk:1974mc,Yoneya:1974jg},  by sigma-model calculations \cite{Callan:1985ia,Fradkin:1984pq,Fradkin:1985fq} or by imposing global symmetries of the string theory in which we are interested.

One of the global symmetries of the  string theory is   T-duality \cite{Giveon:1994fu,Alvarez:1994dn}. This   duality  like  the above gauge symmetries  may be imposed at the action level  to fix the parameters of  the   effective action   at any order of $\alpha'$. One approach for imposing this symmetry  is the   Double Field Theory (DFT) \cite{Siegel:1993xq,Siegel:1993th,Siegel:1993bj,Hull:2009mi,Aldazabal:2013sca} in which the $D$-dimensional  effective action is extended to $2D$-space. In this theory, the gauge transformations are deformed to receive $\alpha'$-corrections whereas  the T-duality symmetry is  imposed without deformation simply by writing the couplings as  $O(D,D)$ scalars \cite{Hohm:2010pp,Aldazabal:2013sca,Hohm:2014xsa,Marques:2015vua,Garousi:2018qes}.  
Another  approach     is  to reduce the $D$-dimensional gauge invariant theory on a circle and impose the T-duality symmetry   by constraining   the couplings in  the $(D-1)$-dimensional spacetime  to be $Z_2$ scalars \cite{Garousi:2017fbe} where $Z_2$-group is the Buscher rules  \cite{Buscher:1987sk,Buscher:1987qj} plus their  $\alpha'$-deformations  \cite{Tseytlin:1991wr,Bergshoeff:1995cg,Kaloper:1997ux}.       Using this approach for the case that $B$-field is zero, the known gravity and dilaton  couplings in the effective actions at orders $\alpha',\alpha'^2,\alpha'^3$ have been found  in   \cite{Razaghian:2017okr,Razaghian:2018svg},  up to some overall factors. Moreover, when $B$-field is non-zero, the known couplings at order $\alpha'$ and their corresponding corrections to the Buscher rules have been found in \cite{Garousi:2019wgz} . In this paper, we are going to use this  approach to find the couplings at order $\alpha'^2$ for the case that $B$-field is non-zero. These couplings,  except its Riemann cubed couplings,  have not been  found by any other methods in string theory. 

It is known  that the effective action at order $\alpha'^2$ depends on the scheme that one uses for the effective action at order $\alpha'$ \cite{Bento1990}. In the T-duality approach,  this is reflected to the T-duality transformations at order $\alpha'$. It has been observed  in \cite{Garousi:2019wgz} that the T-duality transformations at order $\alpha'$ depends on the scheme that one uses for the effective action  at order $\alpha'$. The T-duality transformation corresponding to the effective action at order $\alpha'$  which has only first time derivative \cite{Meissner:1996sa},  is given in \cite{Kaloper:1997ux}.   The T-duality transformations at order $\alpha'$ corresponding to the effective action at order $\alpha'$ in an arbitrary scheme have been found in  \cite{Garousi:2019wgz}. In this paper we are going to  find the effective action at order $\alpha'^2$ that correspond to the   effective action at order $\alpha'$ which has  minimum number of  couplings \cite{Metsaev:1987zx}.

The outline of the paper is as follows: In section 2, we write the known minimum number of couplings at orders   $\alpha'$ and $\alpha'^2$ that the gauge symmetry can fix up to field redefinitions.   In section 3, we impose the T-duality symmetry on the gauge invariant couplings to find their corresponding parameter. The calculations at order   $\alpha'$   have been already done in  \cite{Garousi:2019wgz}. That calculations produce the known couplings in the literature and the corresponding T-duality transformations. The calculations at order $\alpha'^2$ are new. We have found both the effective action and the corresponding  T-duality transformations. However, since the expressions for the T-duality transformations are very lengthy we will   write only the effective action (see \reef{S2f}). We have found that there are only 27 non-zero couplings in the effective action at order $\alpha'^2$. Two of them have  already   been found by the S-matrix calculations \cite{Metsaev:1986yb}. All other terms are new couplings that the T-duality constraint produces.  In section 4, we briefly discuss our results.

  \section{Gauge invariance constraint}\label{sec.2}

The effective action of string theory has a double expansions. One expansion is the genus expansion which includes  the  classical sphere-level and a tower of quantum effects. The other one is a stringy expansion which is an expansion in terms of higher-derivative or   $\alpha'$-expansion. The classical effective action  has the following     $\alpha'$-expansion in the string frame:
\beqa
\bS_{\rm eff}&=&\sum^\infty_{n=0}\alpha'^n\bS_n=\bS_0+\alpha'\bS_1+\alpha'^2 \bS_2+\cdots\ ; \quad \bS_n= -\frac{2}{\kappa^2}\int d^D x\sqrt{-G} e^{-2\Phi}\mathcal{L}_n\labell{seff}
\eeqa
The effective action must be invariant under the coordinate transformations and under the $B$-field   gauge transformations. So the metric $G_{\mu\nu}$, the  antisymmetric  $B$-field and dilaton  $\Phi$      must  appear in the Lagrangian $ \mathcal{L}_n$  trough their field strengths and their covariant derivatives. This requires the effective action at order $\alpha'^0$ to have the following couplings: 
\beqa
\mathcal{L}_0&= &a_1R+a_2\nabla_\alpha\Phi\nabla^\alpha\Phi+a_3H_{\alpha\beta\gamma}H^{\alpha\beta\gamma}\labell{L0}
\eeqa
where $a_1,a_2,a_3$ are three   parameters which can not be fixed by the gauge invariance constraints.

At higher orders of  $\alpha'$, one has the freedom of using   field redefinitions  and the  Bianchi identities. As a result, there are no unique form for the couplings, even the number of couplings are not unique  at the higher orders of $\alpha'$. There are however, schemes in which the number of  couplings are minimum.     It has been shown in     \cite{Metsaev:1987zx} that the minimum number of couplings at order $\alpha'$ is 8. These 8 couplings can  also be written in different schemes. In one particular such  scheme, the couplings can be written as \cite{Garousi:2019cdn}
\beqa
 \mathcal{L}_1&=& \mathcal{L}^1_1+ \mathcal{L}^2_1\labell{L1112}
\eeqa
where $\mathcal{L}^1_1$ includes the minimum number of couplings which do not include the dilaton, \ie,
\beqa
\mathcal{L}^1_1&= &  b_1 R_{\alpha \gamma \beta \delta} R^{\alpha \beta \gamma \delta}
  +b_2 H_{\alpha}{}^{\delta \epsilon} H^{\alpha \beta \gamma} H_{\beta \delta}{}^{\zeta} H_{\gamma \epsilon \zeta} \nn\\
  &&
  + b_3 H_{\alpha \beta}{}^{\delta} H^{\alpha \beta \gamma} H_{\gamma}{}^{\epsilon \zeta} H_{\delta \epsilon \zeta} 
 +b_4 H_{\alpha}{}^{\delta \epsilon} H^{\alpha \beta \gamma} R_{\beta \delta \gamma \epsilon}\labell{L11}
 \eeqa
 and  $\mathcal{L}^2_1$ includes the other couplings  which   all include non-trivially the dilaton, \ie,
\beqa
  \mathcal{L}^2_1&= & b_5 H_{\beta \gamma \delta} H^{\beta \gamma \delta} \nabla_{\alpha}\Phi \nabla^{\alpha}\Phi
 +b_6 H_{\alpha}{}^{\gamma \delta} H_{\beta \gamma \delta} \nabla^{\alpha}\Phi \nabla^{\beta}\Phi \nn\\&&
 +b_7 H_{\alpha}{}^{\gamma \delta} H_{\beta \gamma \delta} \nabla^{\beta}\nabla^{\alpha}\Phi
 + b_8 \nabla_{\alpha}\Phi \nabla^{\alpha}\Phi \nabla_{\beta}\Phi \nabla^{\beta}\Phi \labell{L12}
 \eeqa
 where $b_1,\cdots, b_8$ are eight parameters   which can not be fixed by the gauge invariance  constraints.

At order $\alpha'^2$,   the minimum number of couplings is 60. In one particular minimal scheme in which there is no  $R,\,R_{\mu\nu},\,\nabla_\mu H^{\mu\alpha\beta}$, $ \nabla_\mu\nabla^\mu\Phi$, the couplings are \cite{Garousi:2019cdn}
\begin{eqnarray}
\mathcal{L}_2&=&\mathcal{L}_2^1+\mathcal{L}_2^2 \labell{L2122}
\end{eqnarray}
where $\mathcal{L}_2^1$ has the minimum number of couplings in which the dilaton does not appear, \ie,
\begin{eqnarray}
\mathcal{L}_2^1&=& c_1R_{\alpha}{}^{\epsilon}{}_{\gamma}{}^{\zeta} R^{\alpha \beta \gamma \delta} R_{\beta \zeta \delta \epsilon} + c_2 R_{\alpha \beta}{}^{\epsilon \zeta} R^{\alpha \beta \gamma \delta} R_{\gamma \epsilon \delta \zeta}+c_3 H_{\alpha}{}^{\delta \epsilon} H^{\alpha \beta \gamma} H_{\beta \delta}{}^{\zeta} H_{\gamma}{}^{\iota \kappa} H_{\epsilon \iota}{}^{\mu} H_{\zeta \kappa \mu}\nn\\&& + c_4 H_{\alpha \beta}{}^{\delta} H^{\alpha \beta \gamma} H_{\gamma}{}^{\epsilon \zeta} H_{\delta}{}^{\iota \kappa} H_{\epsilon \zeta}{}^{\mu} H_{\iota \kappa \mu} + c_{5} H_{\alpha \beta}{}^{\delta} H^{\alpha \beta \gamma} H_{\gamma}{}^{\epsilon \zeta} H_{\delta \epsilon}{}^{\iota} H_{\zeta}{}^{\kappa \mu} H_{\iota \kappa \mu}\nn\\&& + c_{6} H_{\alpha}{}^{\delta \epsilon} H^{\alpha \beta \gamma} H_{\beta}{}^{\zeta \iota} H_{\delta \zeta}{}^{\kappa} R_{\gamma \epsilon \iota \kappa}  + c_{7} H_{\alpha}{}^{\delta \epsilon} H^{\alpha \beta \gamma} R_{\beta}{}^{\zeta}{}_{\delta}{}^{\iota} R_{\gamma \zeta \epsilon \iota} + c_{8} H_{\alpha \beta}{}^{\delta} H^{\alpha \beta \gamma} H_{\epsilon \zeta}{}^{\kappa} H^{\epsilon \zeta \iota} R_{\gamma \iota \delta \kappa}\nn\\&& + c_{9} H^{\alpha \beta \gamma} H^{\delta \epsilon \zeta} R_{\alpha \beta \delta}{}^{\iota} R_{\gamma \iota \epsilon \zeta}  + c_{10} H_{\alpha}{}^{\delta \epsilon} H^{\alpha \beta \gamma} R_{\beta}{}^{\zeta}{}_{\delta}{}^{\iota} R_{\gamma \iota \epsilon \zeta} + c_{11} H_{\alpha}{}^{\delta \epsilon} H^{\alpha \beta \gamma} R_{\beta}{}^{\zeta}{}_{\gamma}{}^{\iota} R_{\delta \zeta \epsilon \iota}\nn\\&& + c_{12} H_{\alpha \beta}{}^{\delta} H^{\alpha \beta \gamma} R_{\gamma}{}^{\epsilon \zeta \iota} R_{\delta \zeta \epsilon \iota}  + c_{13} H_{\alpha \beta}{}^{\delta} H^{\alpha \beta \gamma} H_{\gamma}{}^{\epsilon \zeta} H_{\epsilon}{}^{\iota \kappa} R_{\delta \iota \zeta \kappa} + c_{14} H_{\alpha}{}^{\delta \epsilon} H^{\alpha \beta \gamma} H_{\beta \delta}{}^{\zeta} H_{\gamma}{}^{\iota \kappa} R_{\epsilon \iota \zeta \kappa}\nn\\&& + c_{15} H_{\alpha \beta}{}^{\delta} H^{\alpha \beta \gamma} H_{\gamma}{}^{\epsilon \zeta} H_{\delta}{}^{\iota \kappa} R_{\epsilon \iota \zeta \kappa}  + c_{16} H_{\alpha}{}^{\delta \epsilon} H^{\alpha \beta \gamma} \nabla_{\iota}H_{\delta \epsilon \zeta} \nabla^{\iota}H_{\beta \gamma}{}^{\zeta}\nn\\&& + c_{17} H_{\alpha}{}^{\delta \epsilon} H^{\alpha \beta \gamma} \nabla_{\zeta}H_{\gamma \epsilon \iota} \nabla^{\iota}H_{\beta \delta}{}^{\zeta} + c_{18} H_{\alpha}{}^{\delta \epsilon} H^{\alpha \beta \gamma} \nabla_{\iota}H_{\gamma \epsilon \zeta} \nabla^{\iota}H_{\beta \delta}{}^{\zeta}\nn\\&& + c_{19} H_{\alpha \beta}{}^{\delta} H^{\alpha \beta \gamma} \nabla_{\zeta}H_{\delta \epsilon \iota} \nabla^{\iota}H_{\gamma}{}^{\epsilon \zeta} + c_{20} H_{\alpha \beta}{}^{\delta} H^{\alpha \beta \gamma} \nabla_{\iota}H_{\delta \epsilon \zeta} \nabla^{\iota}H_{\gamma}{}^{\epsilon \zeta}\labell{L21}
\end{eqnarray}
 and $\mathcal{L}_2^2$ has  the other couplings which all include  derivatives of the dilaton, \ie,
\begin{eqnarray}
\mathcal{L}_2^2&\!\!\!\!\!=\!\!\!\!\!& c_{21} H_{\beta}{}^{\epsilon \zeta} H^{\beta \gamma \delta} H_{\gamma \epsilon}{}^{\iota} H_{\delta \zeta \iota} \nabla_{\alpha}\Phi \nabla^{\alpha}\Phi + c_{22} R_{\beta \gamma \delta \epsilon} R^{\beta \gamma \delta \epsilon} \nabla_{\alpha}\Phi \nabla^{\alpha}\Phi + c_{23} H_{\beta}{}^{\epsilon \zeta} H^{\beta \gamma \delta} R_{\gamma \epsilon \delta \zeta} \nabla_{\alpha}\Phi \nabla^{\alpha}\Phi\nn\\&& + c_{24} H_{\alpha}{}^{\gamma \delta} H_{\beta}{}^{\epsilon \zeta} H_{\gamma \epsilon}{}^{\iota} H_{\delta \zeta \iota} \nabla^{\alpha}\Phi \nabla^{\beta}\Phi + c_{25} R_{\alpha}{}^{\gamma \delta \epsilon} R_{\beta \delta \gamma \epsilon} \nabla^{\alpha}\Phi \nabla^{\beta}\Phi \nn\\&&+ c_{26} H_{\alpha}{}^{\gamma \delta} H_{\beta}{}^{\epsilon \zeta} R_{\gamma \epsilon \delta \zeta} \nabla^{\alpha}\Phi \nabla^{\beta}\Phi + c_{27} H_{\gamma \delta \epsilon} H^{\gamma \delta \epsilon} \nabla^{\alpha}\Phi \nabla_{\beta}\nabla_{\alpha}\Phi \nabla^{\beta}\Phi \nn\\&&+ c_{28} H_{\alpha}{}^{\gamma \delta} H_{\beta}{}^{\epsilon \zeta} H_{\gamma \epsilon}{}^{\iota} H_{\delta \zeta \iota} \nabla^{\beta}\nabla^{\alpha}\Phi + c_{29} H_{\alpha}{}^{\gamma \delta} H_{\beta}{}^{\epsilon \zeta} H_{\gamma \delta}{}^{\iota} H_{\epsilon \zeta \iota} \nabla^{\beta}\nabla^{\alpha}\Phi \nn\\&&+ c_{30} H_{\alpha}{}^{\gamma \delta} H_{\beta \gamma}{}^{\epsilon} H_{\delta}{}^{\zeta \iota} H_{\epsilon \zeta \iota} \nabla^{\beta}\nabla^{\alpha}\Phi + c_{31} H_{\gamma \delta}{}^{\zeta} H^{\gamma \delta \epsilon} R_{\alpha \epsilon \beta \zeta} \nabla^{\beta}\nabla^{\alpha}\Phi + c_{32} R_{\alpha}{}^{\gamma \delta \epsilon} R_{\beta \delta \gamma \epsilon} \nabla^{\beta}\nabla^{\alpha}\Phi \nn\\&&+ c_{33} H_{\alpha}{}^{\gamma \delta} H_{\gamma}{}^{\epsilon \zeta} R_{\beta \epsilon \delta \zeta} \nabla^{\beta}\nabla^{\alpha}\Phi + c_{34} H_{\alpha}{}^{\gamma \delta} H_{\beta}{}^{\epsilon \zeta} R_{\gamma \epsilon \delta \zeta} \nabla^{\beta}\nabla^{\alpha}\Phi + c_{35} H_{\alpha}{}^{\delta \epsilon} \nabla^{\alpha}\Phi \nabla^{\beta}\Phi \nabla_{\gamma}H_{\beta \delta \epsilon} \nabla^{\gamma}\Phi \nn\\&&+ c_{36} \nabla_{\alpha}\Phi \nabla^{\alpha}\Phi \nabla_{\beta}\Phi \nabla^{\beta}\Phi \nabla_{\gamma}\Phi \nabla^{\gamma}\Phi + c_{37} \nabla_{\alpha}\Phi \nabla^{\alpha}\Phi \nabla^{\beta}\Phi \nabla_{\gamma}\nabla_{\beta}\Phi \nabla^{\gamma}\Phi \nn\\&&+ c_{38} H_{\beta}{}^{\delta \epsilon} H_{\gamma \delta \epsilon} \nabla^{\alpha}\Phi \nabla^{\beta}\Phi \nabla^{\gamma}\nabla_{\alpha}\Phi + c_{39} H_{\beta}{}^{\delta \epsilon} H_{\gamma \delta \epsilon} \nabla^{\beta}\nabla^{\alpha}\Phi \nabla^{\gamma}\nabla_{\alpha}\Phi\nn\\&& + c_{40} \nabla^{\alpha}\Phi \nabla^{\beta}\Phi \nabla_{\gamma}\nabla_{\beta}\Phi \nabla^{\gamma}\nabla_{\alpha}\Phi + c_{41} \nabla^{\beta}\nabla^{\alpha}\Phi \nabla_{\gamma}\nabla_{\beta}\Phi \nabla^{\gamma}\nabla_{\alpha}\Phi \nn\\&&+ c_{42} H_{\beta}{}^{\delta \epsilon} H_{\gamma \delta \epsilon} \nabla_{\alpha}\Phi \nabla^{\alpha}\Phi \nabla^{\gamma}\nabla^{\beta}\Phi + c_{43} H_{\alpha}{}^{\delta \epsilon} \nabla^{\alpha}\Phi \nabla_{\gamma}H_{\beta \delta \epsilon} \nabla^{\gamma}\nabla^{\beta}\Phi \nn\\&&+ c_{44} \nabla_{\alpha}\Phi \nabla^{\alpha}\Phi \nabla_{\gamma}\nabla_{\beta}\Phi \nabla^{\gamma}\nabla^{\beta}\Phi + c_{45} H_{\alpha \gamma}{}^{\epsilon} H_{\beta \delta \epsilon} \nabla^{\alpha}\Phi \nabla^{\beta}\Phi \nabla^{\delta}\nabla^{\gamma}\Phi \nn\\&&+ c_{46} R_{\alpha \gamma \beta \delta} \nabla^{\alpha}\Phi \nabla^{\beta}\Phi \nabla^{\delta}\nabla^{\gamma}\Phi + c_{47} H_{\alpha \gamma}{}^{\epsilon} H_{\beta \delta \epsilon} \nabla^{\beta}\nabla^{\alpha}\Phi \nabla^{\delta}\nabla^{\gamma}\Phi + c_{48} R_{\alpha \gamma \beta \delta} \nabla^{\beta}\nabla^{\alpha}\Phi \nabla^{\delta}\nabla^{\gamma}\Phi \nn\\&&+ c_{49} H_{\beta}{}^{\delta \epsilon} \nabla^{\alpha}\Phi \nabla^{\gamma}\nabla^{\beta}\Phi \nabla_{\epsilon}H_{\alpha \gamma \delta} + c_{50} H^{\gamma \delta \epsilon} \nabla^{\alpha}\Phi \nabla^{\beta}\nabla_{\alpha}\Phi \nabla_{\epsilon}H_{\beta \gamma \delta} \nn\\&&+ c_{51} \nabla^{\alpha}\Phi \nabla^{\beta}\Phi \nabla_{\delta}H_{\beta \gamma \epsilon} \nabla^{\epsilon}H_{\alpha}{}^{\gamma \delta} + c_{52} \nabla^{\beta}\nabla^{\alpha}\Phi \nabla_{\delta}H_{\beta \gamma \epsilon} \nabla^{\epsilon}H_{\alpha}{}^{\gamma \delta} \nn\\&&+ c_{53} \nabla^{\alpha}\Phi \nabla^{\beta}\Phi \nabla_{\epsilon}H_{\beta \gamma \delta} \nabla^{\epsilon}H_{\alpha}{}^{\gamma \delta} + c_{54} \nabla^{\beta}\nabla^{\alpha}\Phi \nabla_{\epsilon}H_{\beta \gamma \delta} \nabla^{\epsilon}H_{\alpha}{}^{\gamma \delta} \nn\\&&+ c_{55} \nabla_{\alpha}\Phi \nabla^{\alpha}\Phi \nabla_{\epsilon}H_{\beta \gamma \delta} \nabla^{\epsilon}H^{\beta \gamma \delta} + c_{56} H_{\alpha}{}^{\beta \gamma} R_{\gamma \zeta \delta \epsilon} \nabla^{\alpha}\Phi \nabla^{\zeta}H_{\beta}{}^{\delta \epsilon} \nn\\&&+ c_{57} H_{\beta \gamma}{}^{\epsilon} H^{\beta \gamma \delta} H_{\delta}{}^{\zeta \iota} \nabla^{\alpha}\Phi \nabla_{\iota}H_{\alpha \epsilon \zeta} + c_{58} H_{\alpha}{}^{\beta \gamma} H_{\delta \epsilon}{}^{\iota} H^{\delta \epsilon \zeta} \nabla^{\alpha}\Phi \nabla_{\iota}H_{\beta \gamma \zeta} \nn\\&&+ c_{59} H_{\alpha}{}^{\beta \gamma} H_{\beta}{}^{\delta \epsilon} H_{\delta}{}^{\zeta \iota} \nabla^{\alpha}\Phi \nabla_{\iota}H_{\gamma \epsilon \zeta} + c_{60} H_{\alpha}{}^{\beta \gamma} H_{\beta}{}^{\delta \epsilon} H_{\gamma}{}^{\zeta \iota} \nabla^{\alpha}\Phi \nabla_{\iota}H_{\delta \epsilon \zeta}\labell{AA}
\end{eqnarray}
 where $c_1,\cdots, c_{60}$ are 60  parameters   which can not be fixed by the gauge invariance  constraint.

Up to this point, the above couplings are valid for any higher derivative  theory which includes metric, $B$-field and dilaton. In the string theory, however, the parameters in \reef{L0}, \reef{L1112} and \reef{L2122} may be fixed by imposing some other specific constraints which are  valid only in the string theory. For example, one may construct the appropriate S-matrix elements with the above couplings and then compare them with the $\alpha'$-expansion of the corresponding sphere-level S-matrix elements in the  string theory to fix the parameters. This method has been used in  \cite{Metsaev:1987zx} to find the parameters in   \reef{L0}, \reef{L1112}. The parameters $c_1,c_2$ in  \reef{L2122} have been also found by the S-matrix method in \cite{Metsaev:1986yb}. The S-matrix  method for fixing all parameters in \reef{L2122}, however,  requires one to calculate six-point function in string theory in full details which has not been done yet. 

Instead of comparing the S-matrix elements of above couplings with the corresponding S-matrix elements in the string theory, one may impose some other symmetries of the string theory   to fix the parameters in \reef{L0}, \reef{L1112} and \reef{L2122} . The bosonic string theory has the global T-duality symmetry as well as the gauge  symmetries that have been used to find the couplings  in \reef{L0}, \reef{L1112} and \reef{L2122}. It has been shown in \cite{Garousi:2019wgz} that the T-duality symmetry can fix correctly  the couplings in \reef{L0}, \reef{L1112} up to   overall factors at each order of $\alpha'$.   In the next section, we show that imposing the T-duality   on the couplings in \reef{L2122} can also  fix all 60 parameters in terms of the overall factor at order $\alpha'$.

 \section{T-duality invariance constraint}\label{sec.3}

 The T-duality  constraint on the $D$-dimensional effective action $\bS_{\rm eff}$,  in the most simple form, is to reduce the theory on a circle with $U(1)$ isometry to find the $(D-1)$-dimensional effective action $S_{\rm eff}(\psi)$ where $\psi$ represents all massless fields in the $(D-1)$-dimensional base space. Then one has to transform it under the T-duality transformations to produce  $S_{\rm eff}(\psi')$ where $\psi'$ represents the T-duality transformations of the massless fields in the base space. The T-duality invariance constraint is then
 \beqa
 S_{\rm eff}(\psi)-S_{\rm eff}(\psi')&=&\int d^{D-1}x \sqrt{-\bg}\nabla_a(e^{-2\bphi}J^a)\labell{TS}
 \eeqa
where $\bg_{ab}$, $\bphi$ are the metric and dilaton in the    base space, and $J^a$ is an arbitrary covariant vector made of the $(D-1)$-dimensional fields. It has the following $\alpha'$-expansion:
\beqa
J^a&=&\sum^\infty_{n=0}\alpha'^nJ^a_n
\eeqa
 where $J^a_n$ is an arbitrary covariant vector at order $\alpha'^n$.

  To have a background with $U(1)$ isometry,  it is convenient to use the following background for  the metric, $B$-field and dilaton:
  \beqa
G_{\mu\nu}=\left(\matrix{\bg_{ab}+e^{\varphi}g_{a }g_{b }& e^{\varphi}g_{a }&\cr e^{\varphi}g_{b }&e^{\varphi}&}\right),\, B_{\mu\nu}= \left(\matrix{\bb_{ab}+\frac{1}{2}b_{a }g_{b }- \frac{1}{2}b_{b }g_{a }&b_{a }\cr - b_{b }&0&}\right),\,  \Phi=\bar{\phi}+\varphi/4\labell{reduc}\eeqa
where $ \bb_{ab}$  is  the   B-field in the base space, and $g_{a},\, b_{b}$ are two vectors  in this space. Inverse of the above $D$-dimensional metric is 
\beqa
G^{\mu\nu}=\left(\matrix{\bg^{ab} &  -g^{a }&\cr -g^{b }&e^{-\varphi}+g_{c}g^{c}&}\right)\labell{inver}
\eeqa
where $\bg^{ab}$ is the inverse of the base  metric which raises the index of the   vectors.

The T-duality transformations at the leading order of $\alpha'$ on the $(D-1)$-dimensional fields   are given by the  Buscher rules \cite{Buscher:1987sk,Buscher:1987qj}. In the above parametrisation, they  become the following linear transformations:
\beqa
\varphi'= -\varphi
\,\,\,,\,\,g'_{\mu }= b_{\mu }\,\,\,,\,\, b'_{\mu }= g_{\mu } \,\,\,,\,\,\bg_{\alpha\beta}'=\bg_{\alpha\beta} \,\,\,,\,\,\bb_{\alpha\beta}'=\bb_{\alpha\beta} \,\,\,,\,\,  \bar{\phi}'= \bar{\phi}\labell{T2}
\eeqa
They form a $Z_2$-group, \ie $ (x')'= x$ where $x$ is any field in the base space. At higher orders of $\alpha'$, the above transformations receive higher derivative corrections, \ie
\beqa
\psi'&=&\sum^\infty_{n=0}\alpha'^n\psi'_n
\eeqa
where $\psi'_0$ is the Buscher rules \reef{T2}, $\psi'_1$ contains corrections to the Buscher rules at order $\alpha'$ and so on. The deformed transformations must satisfy the $Z_2$-group.

\subsection{T-duality constraint at orders $\alpha'^0,\, \alpha'$}

To impose the constraint \reef{TS} on the diffeomorphism invariant couplings  \reef{L2122}, we first review how imposing this constraint  on the couplings  at orders $\alpha'^0$ and $\alpha'$ can  fix their parameters \cite{Garousi:2019wgz}. The constraint \reef{TS} at order $\alpha'^0$ is 
\beqa
S_0(\psi)-S_0(\psi'_0)&=& \int d^{D-1}x \sqrt{-\bg}\nabla_a(e^{-2\bphi}J_0^a)\labell{TS0}
\eeqa
where $J^a_0$ is an arbitrary vector at the leading order of $\alpha'$, and $\psi'_0$ is the Buscher rules \reef{T2}. Reduction of different scalar terms in $\bS_0$ are the following \cite{Garousi:2019wgz}:
\beqa
e^{-2\Phi}\sqrt{-G}&=&e^{-2\bphi}\sqrt{-\bg}\nonumber\\
a_1R&=&a_1(\bR-\nabla^a\nabla_a\vp-\frac{1}{2}\nabla_a\vp \nabla^a\vp -\frac{1}{4}e^{\vp}V^2 )\labell{R}\\
a_2\nabla_{\mu}\Phi\nabla^{\mu}\Phi&=&a_2(\nabla_a\bphi\nabla^a \bphi+\frac{1}{2}\nabla_a\bphi\nabla^a\vp+\frac{1}{16}\nabla_a\vp\nabla^a\vp)\nn\\
a_3H^2&=&a_3(\bH_{abc}\bH^{abc}+3e^{-\vp}W^2) \nn
\eeqa
 where $V_{ab}$ is field strength of the $U(1)$ gauge field $g_{a}$, \ie $V_{ab}=\nabla_{a}g_{b}-\nabla_{b}g_{a}$, and $W_{\mu\nu}$ is field strength of the $U(1)$ gauge field $b_{a}$, \ie $W_{ab}=\nabla_{a}b_{\nu}-\nabla_{b}b_{a}$. The    three-form $\bH$ is defined as $\bH_{abc}=\tilde{H}_{abc}-g_{a}W_{bc}-g_{c}W_{ab}-g_{b}W_{ca}$ where the three-form  $\tilde{H}$ is field strength of the two-form $\bb_{ab}+\frac{1}{2}b_{a}g_{b}-\frac{1}{2}b_b g_a $ in \reef{reduc}.  The three-form $\bH$ is invariant under the Buscher rules and is not the field strength of a two-form. It satisfies the following Bianchi identity  \cite{Kaloper:1997ux}:
 \beqa
\nabla_{[a}\bH_{bcd]}&=&-\frac{3}{2}V_{[ab}W_{cd]}\labell{Bian}
 \eeqa
 which is invariant under the Buscher rules \reef{T2}. 
 
The transformations of different terms in \reef{R} under the Buscher rules \reef{T2} can easily be found. Then the T-duality constraint \reef{TS0} fixes the parameters $a_1,a_2,a_3$   in the $D$-dimensional action \cite{Garousi:2019wgz}, \ie
 \beqa
\bS_0&=& -\frac{2a_1}{\kappa^2}\int d^Dx e^{-2\Phi}\sqrt{-G}\,  \left(  R + 4\nabla_{a}\Phi \nabla^{a}\Phi-\frac{1}{12}H^2\right)\,.\labell{S0bf}
\eeqa
which is the standard effective action at order $\alpha'^0$, up to an overall factor. The overall factor must be $a_1=1$ to be the effective action of string theory. The constraint \reef{TS0} fixes also the form of vector $J_0^a$ in which we are not interested.  

The constraint \reef{TS} at order $\alpha'$ is 
\beqa
S_0(\psi)+\alpha'S_1(\psi)-S_0(\psi'_0+\alpha'\psi'_1)-\alpha'S_1(\psi'_0)&=& \int d^{D-1}x \sqrt{-\bg}\nabla_a[e^{-2\bphi}(J_0^a+\alpha'J_1^a)]\labell{TS1}
\eeqa
where $J^a_1$ is an arbitrary vector at   order of $\alpha'$, and $\psi'_0+\alpha'\psi'_1$ is the Buscher rule plus its  deformation at order $\alpha'$, \ie
\beqa
&&\varphi'= -\varphi+\alpha'\Delta\vp^{(1)}
\,\,\,,\,\,g'_{a }= b_{a }+\alpha'e^{\vp/2}\Delta g^{(1)}_a\,\,\,,\,\, b'_{a }= g_{a }+\alpha'e^{-\vp/2}\Delta b^{(1)}_a \,\,\,,\,\,\nn\\
&&\bg_{ab}'=\bg_{ab}+\alpha'\Delta \bg^{(1)}_{ab} \,\,\,,\,\,\bH_{abc}'=\bH_{abc}+\alpha'\Delta\bH^{(1)}_{abc} \,\,\,,\,\,  \bar{\phi}'= \bar{\phi}+\alpha'\Delta\bphi^{(1)}\labell{T22}
\eeqa
where $\Delta \vp^{(1)}, \cdots,\Delta\bphi^{(1)}$ contains some  contractions of $\nabla\vp,\nabla\bphi, e^{\vp/2}V, e^{-\vp/2}W,\bH,\bR$ and their covariant derivatives at order $\alpha'$. Since the constraint \reef{TS1} should have terms up to order $\alpha'$, in expanding $S_0(\psi'_0+\alpha'\psi'_1)$, one should keep the terms up to order $\alpha'$, \ie
\beqa
S_0(\psi'_0+\alpha'\psi'_1)&=&S_0(\psi'_0)+\alpha'\delta S_0^{(1)}+\cdots
\eeqa
Using this expansion and the constraint \reef{TS0}, one can simplify the constraint \reef{TS1} to the following constraint which is only at order $\alpha'$:
 \beqa
 S_1(\psi)-S_1(\psi'_0)-\delta S_0^{(1)}&=& \int d^{D-1}x \sqrt{-\bg}\nabla_a[e^{-2\bphi} J_1^a]\labell{TS11}
\eeqa
It has been shown in \cite{Garousi:2019wgz} that the above constraint can fix the parameters in the effective action \reef{L1112} as well as the parameters in the corrections to the Buscher rules up to an overall factor. The result is that $\cL_1^2$ is zero and all terms in  $\cL_1^1$ are non-zero, \ie
  \beqa
 \bS_1&=&\frac{-2b_1 }{\kappa^2}\alpha'\int d^{D}x e^{-2\Phi}\sqrt{-G}\Big(   R_{\alpha \beta \gamma \delta} R^{\alpha \beta \gamma \delta} -\frac{1}{2}H_{\alpha}{}^{\delta \epsilon} H^{\alpha \beta \gamma} R_{\beta  \gamma \delta\epsilon}\nn\\
&&\qquad\qquad\qquad\qquad\qquad\quad+\frac{1}{24}H_{\epsilon\delta \zeta}H^{\epsilon}{}_{\alpha}{}^{\beta}H^{\delta}{}_{\beta}{}^{\gamma}H^{\zeta}{}_{\gamma}{}^{\alpha}-\frac{1}{8}H_{\alpha \beta}{}^{\delta} H^{\alpha \beta \gamma} H_{\gamma}{}^{\epsilon \zeta} H_{\delta \epsilon \zeta}\Big)\labell{S1bf}
\eeqa
   Up to the overall factor $b_1$, the above couplings are   the standard effective action of the   string theory   which has been found in \cite{Metsaev:1987zx} by the S-matrix calculations. For the bosonic string theory $b_1=1/4$,   for the heterotic theory  $b_1=1/8$ and for the  superstring theory  $b_1=0$.
   
The corrections to the Buscher rules corresponding to the above action are the following \cite{Garousi:2019wgz}:
\beqa
\Delta \bar{g}_{ab}^{(1)}&=&2b_1\Big(e^\vp V_a {}^c V_{bc}-e^{-\vp}W_a {}^c W_{bc}\Big) \nn\\
\Delta\bphi^{(1)}&=&\frac{b_1}{2}\Big(e^\vp V^2- e^{-\vp}W^2\Big) \nn\\
\Delta\vp^{(1)}&=&2b_1\Big(\nabla_a\vp\nabla^a\vp+e^\vp V^2+e^{-\vp}W^2\Big) \nn\\
\Delta g_{a}^{(1)}&=&b_1\Big(2e^{-\vp/2}\nabla^bW_{ab}+e^{\vp/2}\bH_{abc} V^{bc}-4e^{-\vp/2}\nabla^b\bphi W_{ab}\Big)\nn\\
\Delta b_{a}^{(1)}&=&-b_1\Big(2e^{\vp/2}\nabla^b V_{ab}+e^{-\vp/2}\bH_{abc} W^{b}-4e^{\vp/2}\nabla^b\bphi V_{ab}\Big)\nn\\
\Delta\bH_{abc}^{(1)}&=&12b_1 \nabla_{[a}(W_b{}^d V_{cd]})-3e^{\vp/2} V_{[ab}\Delta g^{(1)}_{c]}-3e^{-\vp/2} W_{[ab}\Delta b _{c]}^{(1)}\labell{dbH1}
\eeqa 
 Replacing these corrections into \reef{T22}, one finds the corresponding T-duality transformations at order $\alpha'$ satisfy the $Z_2$-group as well as the Bianchi identity \reef{Bian} \cite{Garousi:2019wgz}. The constraint \reef{TS11} fixes also the vector $J_1^a$ in which we are not interested.   
 
 \subsection{T-duality constraint at order $  \alpha'^2$}
 
 We now study in details the constraint \reef{TS} at order $\alpha'^2$ to fix the 60 parameters in \reef{L2122}. This constraint at order $\alpha'^2$ is 
 \beqa
&&S_0(\psi)+\alpha'S_1(\psi)+\alpha'^2S_2(\psi)-S_0(\psi'_0+\alpha'\psi'_1+\alpha'^2\psi_2')-\alpha'S_1(\psi'_0+\alpha'\psi_1')-\alpha'^2S_2(\psi_0')\nn\\
 &&\qquad\qquad\qquad\qquad\qquad\qquad= \int d^{D-1}x \sqrt{-\bg}\nabla_a[e^{-2\bphi}(J_0^a+\alpha'J_1^a+\alpha'^2 J_2^a)]\labell{TS2}
\eeqa
where $J^a_2$ is an arbitrary vector at   order  $\alpha'^2$,  $ \alpha'\psi'_1$ represents  the   corrections to the Buscher rules at order $\alpha'$, \eg \reef{dbH1},  and  $\alpha'^2\psi_2'$ represents the corrections to the Buscher rules at order $\alpha'^2$, \ie 
\beqa
&&\varphi'= -\varphi+\alpha'\Delta\vp^{(1)}+\frac{1}{2}\alpha'^2\Delta\vp^{(2)}
\,\,\,,\,\,g'_{a }= b_{a }+\alpha'e^{\vp/2}\Delta g^{(1)}_a+\frac{1}{2}\alpha'^2e^{\vp/2}\Delta g^{(2)}_a \nn\\
&& b'_{a }= g_{a }+\alpha'e^{-\vp/2}\Delta b^{(1)}_a +\frac{1}{2}\alpha'^2e^{-\vp/2}\Delta b^{(2)}_a\,\,\,,\,\,\bg_{ab}'=\bg_{ab}+\alpha'\Delta \bg^{(1)}_{ab}+\frac{1}{2}\alpha'^2\Delta \bg^{(2)}_{ab}\nn\\ &&\bH_{abc}'=\bH_{abc}+\alpha'\Delta\bH^{(1)}_{abc}+\frac{1}{2}\alpha'^2\Delta\bH^{(2)}_{abc} \,\,\,,\,\,  \bar{\phi}'= \bar{\phi}+\alpha'\Delta\bphi^{(1)}+\frac{1}{2}\alpha'^2\Delta\bphi^{(2)}\labell{T221}
\eeqa
where $\Delta \vp^{(2)}, \cdots,\Delta\bphi^{(2)}$ contains all  contractions of $\nabla\vp,\nabla\bphi, e^{\vp/2}V, e^{-\vp/2}W,\bH,\bR$ and their covariant derivatives at order $\alpha'^2$ with  unknown coefficients. Since the constraint \reef{TS2} should have terms up to order $\alpha'^2$, in expanding $S_0(\psi'_0+\alpha'\psi'_1+\alpha'^2\psi'_2)$ and $\alpha'S_1(\psi'_0+\alpha'\psi'_1 )$, one should keep the terms up to order $\alpha'^2$, \ie
\beqa
S_0(\psi'_0+\alpha'\psi'_1+\alpha'^2\psi'_2)&=&S_0(\psi'_0)+\alpha'\delta S_0^{(1)}+\alpha'^2\delta S_0^{(2)} +\cdots\nn\\
\alpha'S_1(\psi'_0+\alpha'\psi'_1 )&=&\alpha'S_1(\psi'_0)+\alpha'^2\delta S_1^{(1)}+\cdots 
\eeqa
Using these expansions and the constraints \reef{TS0} and \reef{TS11}, one can simplify the constraint \reef{TS2} to the following constraint which is only at order $\alpha'^2$:
 \beqa
 S_2(\psi)-S_2(\psi'_0)-\delta S_0^{(2)}-\delta S_1^{(1)}&=& \int d^{D-1}x \sqrt{-\bg}\nabla_a[e^{-2\bphi} J_2^a]\labell{TS21}
\eeqa
The speculation is that this constraint as well as the constraint that the T-duality transformations should satisfy the $Z_2$-group and the anomalous Bianchi identity \reef{Bian} may fix all parameters in the diffiomorphism invariant couplings \reef{L2122}. 

The T-duality transformations \reef{T221} should satisfy the $Z_2$-group. This produces the following constraint between the corrections at orders $\alpha'$ and $\alpha'^2$:
\beqa
-\Delta\vp^{(2)}(\psi)+\Delta\vp^{(2)}(\psi_0')+2\delta\Delta\vp^{(1)}(\psi_0')&=&0\nn\\
\Delta b_a^{(2)}(\psi)+\Delta g_a^{(2)}(\psi'_0)+2\delta\Delta g_a^{(1)}(\psi_0')&=&0\nn\\
\Delta g_a^{(2)}(\psi)+\Delta b_a^{(2)}(\psi'_0)+2\delta\Delta b_a^{(1)}(\psi_0')&=&0\nn\\
\Delta \bg_{ab}^{(2)}(\psi)+\Delta \bg_{ab}^{(2)}(\psi'_0)+2\delta\Delta \bg_{ab}^{(1)}(\psi_0')&=&0\nn\\
\Delta \bH_{abc}^{(2)}(\psi)+\Delta \bH_{abc}^{(2)}(\psi'_0)+2\delta\Delta \bH_{abc}^{(1)}(\psi_0')&=&0\nn\\
\Delta\bphi^{(2)}(\psi)+\Delta\bphi^{(2)}(\psi_0')+2\delta\Delta\bphi^{(1)}(\psi_0')&=&0\labell{Z2}
\eeqa
where we have used the $Z_2$-constraint at order $\alpha'$ which are \cite{Garousi:2019wgz} 
\beqa
-\Delta\vp^{(1)}(\psi)+\Delta\vp^{(1)}(\psi_0') &=&0\nn\\
\Delta b_a^{(1)}(\psi)+\Delta g_a^{(1)}(\psi'_0) &=&0\nn\\
\Delta g_a^{(1)}(\psi)+\Delta b_a^{(1)}(\psi'_0)&=&0\nn\\
\Delta \bg_{ab}^{(1)}(\psi)+\Delta \bg_{ab}^{(1)}(\psi'_0)&=&0\nn\\
\Delta \bH_{abc}^{(1)}(\psi)+\Delta \bH_{abc}^{(1)}(\psi'_0) &=&0\nn\\
\Delta\bphi^{(1)}(\psi)+\Delta\bphi^{(1)}(\psi_0') &=&0\labell{Z22}
\eeqa
The perturbations $\delta\Delta\vp^{(1)}(\psi_0'), \cdots,\delta\Delta\bphi^{(1)}(\psi_0') $ in \reef{Z2} are defined as
\beqa
\Delta\vp^{(1)}(\psi_0'+\alpha'\psi_1')&=&\Delta\vp^{(1)}(\psi_0')+\alpha'\delta\Delta\vp^{(1)}(\psi_0')+\cdots\nn\\
e^{\vp/2}(\psi_0'+\alpha'\psi_1')\Delta g_a^{(1)}(\psi_0'+\alpha'\psi_1')&=&e^{-\vp/2}\Delta g_a^{(1)}(\psi_0' )+\alpha'e^{-\vp/2}\delta\Delta g_a^{(1)}(\psi_0')+\cdots\nn\\
e^{-\vp/2}(\psi_0'+\alpha'\psi_1')\Delta b_a^{(1)}(\psi_0'+\alpha'\psi_1')&=&e^{\vp/2}\Delta b_a^{(1)}(\psi_0' )+\alpha'e^{\vp/2}\delta\Delta b_a^{(1)}(\psi_0')+\cdots\nn\\
\Delta\bg_{ab}^{(1)}(\psi_0'+\alpha'\psi_1')&=&\Delta\bg_{ab}^{(1)}(\psi_0' )+\alpha'\delta\Delta \bg_{ab}^{(1)}(\psi_0')+\cdots\nn\\
\Delta \bH_{abc}^{(1)}(\psi_0'+\alpha'\psi_1')&=&\Delta \bH_{abc}^{(1)}(\psi_0')+\alpha'\delta\Delta \bH_{abc}^{(1)}(\psi_0')+\cdots\nn\\
\Delta\bphi^{(1)}(\psi_0'+\alpha'\psi_1')&=&\Delta\bphi^{(1)}(\psi_0' )+\alpha'\delta\Delta\bphi^{(1)}(\psi_0')+\cdots
\eeqa
where dots represent the perturbations at higher orders of $\alpha'$ which do not appear in our calculations.

The Bianchi identity \reef{Bian} in   terms of 3-form $H$ and 1-forms $g$, $b$  is $dH=-(3/2)dg\wedge db$. The T-dual fields should satisfy this identity as well, \ie
\beqa
d(\bH+\alpha'\Delta\bH^{(1)}+\frac{1}{2}\alpha'^2\Delta\bH^{(2)}+\cdots)&=&-\frac{3}{2}d(b+\alpha'e^{\vp/2}\Delta g^{(1)}+\frac{1}{2}\alpha'^2e^{\vp/2}\Delta g^{(2)}+\cdots)\nn\\
&&\wedge d( g+\alpha'e^{-\vp/2}\Delta b^{(1)}  +\frac{1}{2}\alpha'^2e^{-\vp/2}\Delta b^{(2)} +\cdots)\labell{aa}
\eeqa
This relation at order $\alpha'^0$ gives the Bianchi identity \reef{Bian}. At order $\alpha'$ it gives the following relation between the corrections to the Buscher rules at order $\alpha'$:
\beqa
\Delta\bH^{(1)}&=&\tilde H^{(1)}-\frac{3}{2}\Big[db\wedge (e^{-\vp/2}\Delta b^{(1)})+(e^{\vp/2}\Delta g^{(1)})\wedge dg\Big]
\eeqa
where $\tilde H^{(1)}$ is a $U(1)\times U(1)$ invariant closed 3-form, \ie $d \tilde H^{(1)}=0$, at order $\alpha'$ which is odd under parity. The corrections \reef{dbH1} satisfy this relation. At order $\alpha'^2$, the Bianchi identity \reef{aa} produces the following relation between the corrections at orders $\alpha'$ and $\alpha'^2$:
\beqa
\Delta\bH^{(2)}&=&\tilde H^{(2)}-\frac{3}{2}\Big[db\wedge (e^{-\vp/2}\Delta b^{(2)})+(e^{\vp/2}\Delta g^{(2)})\wedge dg\nn\\
&&+(e^{\vp/2}\Delta g^{(1)})\wedge d(e^{-\vp/2}\Delta b^{(1)})+d(e^{\vp/2}\Delta g^{(1)})\wedge (e^{-\vp/2}\Delta b^{(1)})\Big]\labell{HH2}
\eeqa
where $\tilde H^{(2)}$ is a   closed 3-form, \ie $d\tilde H^{(2)}=0$,   which  contains all  contractions of $\nabla\vp$, $\nabla\bphi$,  $e^{\vp/2}V, e^{-\vp/2}W$, $\bH,\bR$ and their covariant derivatives at order $\alpha'^2$ with  unknown coefficients.

Therefore, the second order corrections $\Delta\vp^{(2)},\, \Delta g_a^{(2)},\, \Delta b_a^{(2)}, \,\Delta \bg_{ab}^{(2)}$ and $\Delta\bphi^{(2)}$ should be  all  contractions of $\nabla\vp,\nabla\bphi, e^{\vp/2}V, e^{-\vp/2}W,\bH,\bR$ and their covariant derivatives at order $\alpha'^2$ with  unknown coefficients. The correction $\Delta \bH_{abc}^{(2)}$ is then can be calculated from \reef{HH2}. All corrections should satisfy the $Z_2$-relations \reef{Z2}. They produce some algebraic equations between the parameters of the corrections at order $\alpha'^2$ and the parameter $b_1$ in the corrections at order $\alpha'$, \ie \reef{dbH1}. These parameters and the 60 parameters in the action \reef{L2122} should satisfy the constraint \reef{TS21} as well.

To use the constraint \reef{TS21} one needs to reduce the couplings in \reef{L2122}. The reduction of each term at order $\alpha'^2$ is a very lengthy expression. However, the final result for the reduction of each term must be an invariant term under the  $U(1)\times U(1)$ gauge transformations. Using this fact as a constraint, the calculations of the reduction of $\bS_2$ can be simplified greatly. The couplings in \reef{L2122} have only Riemann curvature, $H$, $\nabla H$, $\nabla\Phi$ and $\nabla\nabla\Phi$. So we need to reduce these terms and then contract them with the metric \reef{inver}. In the  reduction of these terms, there are many terms which contains gauge field  $g_a$ without its field strength. These terms must be cancelled at the end of the day for the scalar couplings. Hence, to simplify the calculation we drop those terms in the reduction of  $R_{\mu\nu\alpha\beta}$, $H_{\mu\nu\alpha}$, $\nabla_{\mu} H_{\nu\alpha\beta}$, $\nabla_{\mu}\Phi$, $\nabla_{\mu}\nabla_{\nu}\Phi$  and $G^{\mu\nu}$ which have the gauge field $g_a$.  Using this simplification, the reduction of Riemann curvature becomes\footnote{We have used the package ''xAct" \cite{Nutma:2013zea} for performing the calculations in this paper.}
\beqa
R_{abcd}&=&\bR_{abcd}+\frac{1}{4}e^\vp (V_{ad}V_{bc}-V_{ac}V_{bd}-2V_{ab}V_{cd})\nn\\
R_{abcy}&=&\frac{1}{4}e^\vp(V_{bc}\nabla_a\vp-V_{ac}\nabla_b\vp-2V_{ab}\nabla_c\vp-2\nabla_cV_{ab})\nn\\
R_{aycy}&=&\frac{1}{4}e^\vp(e^\vp V_a{}^bV_{cb}-\nabla_a\vp\nabla_c\vp-2\nabla_c\nabla_a\vp)\labell{rR}
\eeqa
All other components  are either zero or related to the above terms by the Riemann symmetries.  The reduction of different components of $\nabla\nabla\Phi$ and  $\nabla\Phi$ become
\beqa
\nabla_a\nabla_b\Phi&=&\nabla_a\nabla_b\bphi+\frac{1}{4}\nabla_a\nabla_b\vp\nn\\
\nabla_a\nabla_y\Phi&=&-\frac{1}{2}e^\vp(V_{ab}\nabla^b\bphi+\frac{1}{4}V_{ab}\nabla^b\vp)\nn\\
\nabla_y\nabla_y\Phi&=&\frac{1}{2}e^\vp(\nabla_a\vp\nabla^a\bphi+\frac{1}{4}\nabla_a\vp\nabla^a\vp)\nn\\
 \nabla_a\Phi&=&\nabla_a\bphi+\frac{1}{4}\nabla_a\vp\,\,\,;\,\,\,
\nabla_y\Phi\,=\,0\labell{rP}
\eeqa
The reduction of different components of  $\nabla H$ and  $H$ become
\beqa
\nabla_aH_{bcd}&=&\frac{1}{2}(V_{ad}W_{bc}-V_{ac}W_{bd}+V_{ab}W_{cd}+2\nabla_a\bH_{bcd})\nn\\
\nabla_aH_{bcy}&=&\frac{1}{2}(-e^\vp\bH_{bcd}V_{a}{}^d-W_{bc}\nabla_a\vp+2\nabla_a W_{bc})\nn\\
\nabla_y H_{bcd}&=&\frac{1}{2}e^\vp(\bH_{bda}V_c{}^a-\bH_{cda}V_b{}^a-\bH_{bca}V_d{}^a)+\frac{1}{2}(W_{bd}\nabla_c\vp-W_{cd}\nabla_b\vp-W_{bc}\nabla_d\vp)\nn\\
\nabla_yH_{bcy}&=&\frac{1}{2}e^\vp(\bH_{bca}\nabla^a\vp-V_c{}^aW_{ba}+V_b{}^aW_{ca})\labell{rDH}\\
 H_{abc}&=&\bH_{abc}\,\,\,;\,\,\, H_{aby}\,\,=\,\,W_{ab}\nn
 \eeqa
 The covariant derivatives on the right-hand side of \reef{rR}, \reef{rP} and \reef{rDH} are $(D-1)$-dimensional, and the indices are raised by the inverse metric $\bg^{ab}$. The reduction of inverse of the $D$-dimensional metric in this case also becomes
 \beqa
G^{\mu\nu}=\left(\matrix{\bg^{ab} & 0&\cr 0&e^{-\varphi}&}\right)\labell{inver1}
\eeqa
Using above reductions, one can calculate the reduction of different scalar terms in \reef{L2122}. 

Then using the constraint \reef{TS21}, one finds some   equations involving the 60 parameters in \reef{L2122}, the arbitrary parameters of $J_2^a$ and  the parameters of  $\Delta\vp^{(2)},\, \Delta g_a^{(2)},\, \Delta b_a^{(2)}, \,\Delta \bg_{ab}^{(2)}$, $\Delta\bphi^{(2)}$ and $\tilde{H}^{(2)}$. The parameters  of $\Delta\vp^{(2)},\, \Delta g_a^{(2)},\, \Delta b_a^{(2)}, \,\Delta \bg_{ab}^{(2)}$, $\Delta\bphi^{(2)}$ and $\tilde{H}^{(2)}$ should also satisfy the constraints \reef{Z2} and \reef{HH2}. To solve these  constraints, one has to write the couplings in them in terms of independent couplings. Then   coefficients  of the independent couplings which involve the above parameters should be zero.  To perform this last steps, one has to impose  the following Bianchi identities into the constraints as well:
\beqa
 \bR_{a[bcd]}&=&0,\nn\\
 \nabla_{[a}\bR_{bc]de}&=&0\nn \\
{[}\nabla,\nabla {]}\cO &=& \bR\cO\nn\\
\nabla_{[a}V_{bc]}&=&0\nn\\
\nabla_{[a}W_{bc]}&=&0\nn\\
\nabla_{[a}\bH_{bcd]}+\frac{3}{2}V_{[ab}W_{cd]}&=&0\labell{iden}
\eeqa
To impose the last identity above, we     contract it with tensors  $\nabla\vp$, $\nabla\bphi$,  $e^{\vp/2}V, e^{-\vp/2}W$, $\bH,\bR$  and their derivatives with arbitrary parameters and then add them to the constraints.  To impose the first three identities above,   we  use the locally inertial frame in which these   identities are    automatically satisfied. 
In the locally inertial frame, the metric $\bg_{ab}$ takes its canonical form and its first derivatives are all vanish, \ie,
\[
\bg_{ab}=\eta_{ab},\qquad \partial_a \bg_{bc}=0
\]
The second and higher derivatives of metric, however, are non-zero. In this coordinate,   by rewriting the covariant derivative in terms of partial derivatives, one finds the first three identities in \reef{iden}  are satisfied. To satisfy  the Bianchi identities $dV=0=dW$ as well, in the couplings which involve derivatives of $V$ and $W$, we   rewrite  them    in terms of  their gauge fields, \ie  $V_{ab}=\partial_a g_{b}-\partial_b g_{a} $ and  $W_{ab}=\partial_a b_{b}-\partial_b b_{a} $.  

After using the above steps to write the couplings in the constraints \reef{TS21}, \reef{Z2} and \reef{HH2} in terms of  independent couplings in the local frame, one can set their coefficients to zero to produce some algebraic equations involving only the   parameters. Interestingly, these algebraic equations    fix all the 60 parameters in \reef{L2122} in terms of $b_1$, the overall factor at order $\alpha'$ which should be $b_1=1/4$ for the bosonic string theory. All 20 parameters in \reef{L21} are non-zero and only 7 parameters in \reef{AA} are non-zero. They are  
\beqa
\bS_2&=&\frac{-2b_1 }{\kappa^2}\alpha'^2\int d^{D}x e^{-2\Phi}\sqrt{-G}\Big(  - \frac{1}{12} H_{\alpha}{}^{\delta \epsilon} H^{\alpha \beta 
\gamma} H_{\beta \delta}{}^{\zeta} H_{\gamma}{}^{\iota \kappa} 
H_{\epsilon \iota}{}^{\mu} H_{\zeta \kappa \mu}\nn\\&& + 
\frac{1}{30} H_{\alpha \beta}{}^{\delta} H^{\alpha \beta 
\gamma} H_{\gamma}{}^{\epsilon \zeta} H_{\delta}{}^{\iota 
\kappa} H_{\epsilon \zeta}{}^{\mu} H_{\iota \kappa \mu} + 
\frac{3}{10} H_{\alpha \beta}{}^{\delta} H^{\alpha \beta 
\gamma} H_{\gamma}{}^{\epsilon \zeta} H_{\delta 
\epsilon}{}^{\iota} H_{\zeta}{}^{\kappa \mu} H_{\iota \kappa 
\mu} \nn\\&&+ \frac{13}{20} H_{\alpha}{}^{\epsilon \zeta} 
H_{\beta}{}^{\iota \kappa} H_{\gamma \epsilon \zeta} H_{\delta 
\iota \kappa} R^{\alpha \beta \gamma \delta} + \frac{2}{5} 
H_{\alpha}{}^{\epsilon \zeta} H_{\beta \epsilon}{}^{\iota} 
H_{\gamma \zeta}{}^{\kappa} H_{\delta \iota \kappa} R^{\alpha 
\beta \gamma \delta}\nn\\&& + \frac{18}{5} H_{\alpha 
\gamma}{}^{\epsilon} H_{\beta}{}^{\zeta \iota} H_{\delta 
\zeta}{}^{\kappa} H_{\epsilon \iota \kappa} R^{\alpha \beta 
\gamma \delta} -  \frac{43}{5} H_{\alpha \gamma}{}^{\epsilon} 
H_{\beta}{}^{\zeta \iota} H_{\delta \epsilon}{}^{\kappa} 
H_{\zeta \iota \kappa} R^{\alpha \beta \gamma \delta} \nn\\&&-  
\frac{16}{5} H_{\alpha \gamma}{}^{\epsilon} H_{\beta 
\delta}{}^{\zeta} H_{\epsilon}{}^{\iota \kappa} H_{\zeta \iota 
\kappa} R^{\alpha \beta \gamma \delta} - 2 H_{\beta 
\epsilon}{}^{\iota} H_{\delta \zeta \iota} 
R_{\alpha}{}^{\epsilon}{}_{\gamma}{}^{\zeta} R^{\alpha \beta 
\gamma \delta} - 2 H_{\beta \delta}{}^{\iota} H_{\epsilon 
\zeta \iota} R_{\alpha}{}^{\epsilon}{}_{\gamma}{}^{\zeta} 
R^{\alpha \beta \gamma \delta}\nn\\&& -  \frac{4}{3} 
R_{\alpha}{}^{\epsilon}{}_{\gamma}{}^{\zeta} R^{\alpha \beta 
\gamma \delta} R_{\beta \zeta \delta \epsilon} + \frac{4}{3} 
R_{\alpha \beta}{}^{\epsilon \zeta} R^{\alpha \beta \gamma 
\delta} R_{\gamma \epsilon \delta \zeta} + 3 
H_{\beta}{}^{\zeta \iota} H_{\epsilon \zeta \iota} R^{\alpha 
\beta \gamma \delta} R_{\gamma}{}^{\epsilon}{}_{\alpha \delta} 
\nn\\&&+ 2 H_{\beta \epsilon}{}^{\iota} H_{\delta \zeta \iota} 
R^{\alpha \beta \gamma \delta} 
R_{\gamma}{}^{\epsilon}{}_{\alpha}{}^{\zeta} + 2 H_{\alpha 
\beta \epsilon} H_{\delta \zeta \iota} R^{\alpha \beta \gamma 
\delta} R_{\gamma}{}^{\epsilon \zeta \iota} + \frac{13}{10} 
H_{\alpha}{}^{\gamma \delta} H_{\beta \gamma}{}^{\epsilon} 
H_{\delta}{}^{\zeta \iota} H_{\epsilon \zeta \iota} 
\nabla^{\beta}\nabla^{\alpha}\Phi\nn\\&& + \frac{13}{5} 
H_{\gamma}{}^{\epsilon \zeta} H_{\delta \epsilon \zeta} 
R_{\alpha}{}^{\gamma}{}_{\beta}{}^{\delta} 
\nabla^{\beta}\nabla^{\alpha}\Phi -  \frac{52}{5} H_{\beta 
\delta}{}^{\zeta} H_{\gamma \epsilon \zeta} 
R_{\alpha}{}^{\gamma \delta \epsilon} 
\nabla^{\beta}\nabla^{\alpha}\Phi\nn\\&& -  \frac{26}{5} H_{\alpha 
\gamma \epsilon} H_{\beta \delta \zeta} R^{\gamma \delta 
\epsilon \zeta} \nabla^{\beta}\nabla^{\alpha}\Phi + 
\frac{13}{5} \nabla^{\beta}\nabla^{\alpha}\Phi 
\nabla_{\epsilon}H_{\beta \gamma \delta} 
\nabla^{\epsilon}H_{\alpha}{}^{\gamma \delta} \nn\\&&+ \frac{13}{10} 
H_{\beta \gamma}{}^{\epsilon} H^{\beta \gamma \delta} 
H_{\delta}{}^{\zeta \iota} \nabla^{\alpha}\Phi 
\nabla_{\iota}H_{\alpha \epsilon \zeta}  -  \frac{13}{20} 
H_{\alpha}{}^{\beta \gamma} H_{\delta \epsilon}{}^{\iota} 
H^{\delta \epsilon \zeta} \nabla^{\alpha}\Phi 
\nabla_{\iota}H_{\beta \gamma \zeta} \nn\\&&+ \frac{1}{20} 
H_{\alpha}{}^{\delta \epsilon} H^{\alpha \beta \gamma} 
\nabla_{\iota}H_{\delta \epsilon \zeta} 
\nabla^{\iota}H_{\beta \gamma}{}^{\zeta} + \frac{1}{5} 
H_{\alpha}{}^{\delta \epsilon} H^{\alpha \beta \gamma} 
\nabla_{\zeta}H_{\gamma \epsilon \iota} 
\nabla^{\iota}H_{\beta \delta}{}^{\zeta}\nn\\&& -  \frac{6}{5} 
H_{\alpha}{}^{\delta \epsilon} H^{\alpha \beta \gamma} 
\nabla_{\iota}H_{\gamma \epsilon \zeta} 
\nabla^{\iota}H_{\beta \delta}{}^{\zeta} -  \frac{6}{5} 
H_{\alpha \beta}{}^{\delta} H^{\alpha \beta \gamma} 
\nabla_{\zeta}H_{\delta \epsilon \iota} 
\nabla^{\iota}H_{\gamma}{}^{\epsilon \zeta}\nn\\&&\qquad\qquad\qquad\qquad\qquad\qquad\qquad  + \frac{17}{10} 
H_{\alpha \beta}{}^{\delta} H^{\alpha \beta \gamma} 
\nabla_{\iota}H_{\delta \epsilon \zeta} 
\nabla^{\iota}H_{\gamma}{}^{\epsilon \zeta}\Big)\labell{S2f}
\eeqa
Note that the 60 parameters are fixed when the Bianchi constraint \reef{HH2} is imposed as well as the T-duality constraint \reef{TS21} and \reef{Z2}. If  one only uses the constraint   \reef{TS21} and \reef{Z2}, then 12 parameters of \reef{L2122} remain arbitrary. It is the constraint  \reef{HH2} which fixes these 12 parameters as well.

The algebraic equations also fix  some of the parameters in the T-duality transformations and the parameters of total derivative terms at order $\alpha'^2$  in terms of $b_1$, and leave many of them to be arbitrary. Some of the  arbitrary parameters in the T-duality transformations may be removed by the Bianchi identities and  some of them are related to the coordinate transformations at order $\alpha'^2$. Even when all the arbitrary parameters are set to zero, there are still too may terms in the T-duality transformations at order $\alpha'^2$, so we do not   write them explicitly. On the other hand, those corrections are only needed if one would like to extend the above couplings to the order $\alpha'^3$ in the bosonic theory in which we are not interested in this paper. The important part of the calculations is that there are 60 relations between the  60 parameters in \reef{L2122} and the parameter $b_1$, \ie the T-duality constraint fixes  all 60 parameters at order $\alpha'^2$ in terms of the overall factor of the couplings at order $\alpha'$!  This ends our illustration of the fact that the T-duality constraint on the effective action can fix uniquely  the effective action of bosonic string  theory at order $\alpha'^2$. 

\section{Discussion}

In this paper, we have shown that  imposing the gauge symmetries and the T-duality symmetry on the effective action of string theory for metric, $B$-field and dilaton at order $\alpha'^2$, can fix the effective action, \ie \reef{S2f}, up to an overall factor which is the overall factor of the effective action at order $\alpha'$. This is extension of the similar calculation at order $\alpha'$  done in \cite{Garousi:2019wgz} which fixes the effective action at order $\alpha'$ up to the  overall factor $b_1$, \ie \reef{S1bf}. In fact, the gauge symmetries require to have 60 couplings at order $\alpha'^2$ with unfixed coefficients \cite{Garousi:2019cdn}, and the T-duality symmetry which is imposed  on the reduction of the effective action   on a circle,  fixes these 60 parameters. 

In the base space, we have done the calculations   in  the local frame in which the first derivatives of the base metric is zero. After solving the constraints, we have  imposed the solution for the parameters in the constraints \reef{TS21}, \reef{Z2}, \reef{HH2}  and found that they are satisfied even when the first derivative of metric is non-zero. It is as expected, because the constraints are some   covariant  identities. If they satisfy in one particular frame like the local inertial frame, they would  satisfy in all other frames as well. 

Most of the couplings in \reef{S2f} are new couplings which have not been found in the literature by other methods in string theory. When $B$-field is zero, the   couplings \reef{S2f}  reduce to two Riemann cubed terms that their coefficients, after using the cyclic symmetry of the Riemann curvature,  become exactly the same as the coefficients  that have been found in  \cite{Metsaev:1986yb} by the S-matrix method. These couplings are invariant under the field redefinitions. However, the couplings which have $B$-field are not invariant under the field redefinitions. When $B$-field is non-zero, one may check the couplings involving four fields with the corresponding four-point S-matrix elements in bosonic  string theory. To check this comparison,    one has  to use a field redefinition that change the Riemann squared terms in \reef{S1bf} to the Gauss-Bonnet  combination in which the propagators do not receive $\alpha'$-correction.  That field redefinitions  would then change the form of the couplings in \reef{S2f}. The resulting couplings then may be checked with the corresponding S-matrix elements. We leave the details of this calculation for the future works.

We have found that 7  dilaton couplings in \reef{AA} are non-zero. On the other hand, it is known that  the couplings at order $\alpha'^2$ depends on the effective action at order $\alpha'$  \cite{Bento1990}. We have used the minimal action \reef{S1bf} and the corresponding T-duality transformations \reef{dbH1}. Using another  scheme for the couplings at order $\alpha'$, some of  the parameters in \reef{S2f} may be changed. It would be interesting to check if there is a scheme for the couplings at order $\alpha'$ for which all the dilaton couplings in \reef{AA} become zero.  

We have found the effective action \reef{S2f} by imposing only the symmetries of string theory, \ie the $B$-field gauge invariance, diffeomorphism  and T-duality invariances. As a result, the effective action   \reef{S2f} is background independent. However, the total derivative terms are   ignored 
 in imposing the T-duality constraint. Hence the effective action \reef{S2f} is valid for all backgrounds that have no boundary. It would be interesting to take into account in details  the total derivative terms to find the boundary terms as well as the bulk terms for the general case that the background has    boundary.   

We have done the calculations in the curved   base space to make sure that the constraints  \reef{TS21}, \reef{Z2}, \reef{HH2} are satisfied in full details. We have performed the calculations in flat base space as well and found exactly the same parameters for \reef{L2122} as in \reef{S2f}. In the  T-duality calculations at order $\alpha'$    \cite{Garousi:2019wgz} which have correctly reproduced the effective action at order $\alpha'$, it is also assumed that the base space is flat. Hence, for the calculations at the higher orders of $\alpha'$ which would be very lengthy calculations, one may safely assume the base space is flat. The most simple calculations at   order $\alpha'^3$  is for superstring theory in which the T-duality transformations have no deformation at orders $\alpha'$ and $\alpha'^2$.  It would be interesting to perform this calculations at order $\alpha'^3$ in the superstring to find the $B$-field couplings which are not known in the literature. 

 If one extends the calculations in the  bosonic theory to the   order  $\alpha'^3$, one would find a   set of couplings   which are proportional to  $b_1$ and another   set of couplings that their overall factor is arbitrary. The comparison with the four-point S-matrix elements  dictates that this factor should be $\zeta(3)$.  At order $\alpha'^4$, again one should find a   set of couplings which are proportional to $b_1$, a set of couplings proportial to $\zeta(3)$ and some other   sets of couplings that their overall factor may be fixed by the corresponding S-matrix elements. Continuing these logic, one would find   sets of couplings at each order of $\alpha'$ which are proportional to $b_1$. Hence, one expects the T-duality constraint produces a   set of couplings at each order of $\alpha'$ that are proportional to $b_1$.  They form a complete set of couplings which would be  invariant   under the T-duality transformations at all orders of $\alpha'$. That T-dual set of  couplings may have de Sitter solution  \cite{Hohm:2019jgu}. It would be interesting to find this T-dual set.
 
 In this paper, while we have deformed the T-duality transformations, we have assumed the gauge transformations are the standard diffeomorphisms and $B$-field gauge transformations which are the correct transformations in the bosonic and superstring theories. In the superstring theory $b_1=0$, hence,  the   couplings \reef{S2f} are zero in the superstring theory as expected.
On the other hand, the 60 parameters in \reef{S2f} do not dependent on the dimension of spacetime. That does not indicate   the   result \reef{S2f} is valid also for the heterotic theory for $b_1=1/8$. The reason is that in the heterotic theory the $B$-field gauge transformation is deformed at order $\alpha'$ which is resulted from the Green-Schwarz anomaly cancellation mechanism  \cite{Green:1984sg}. To produce the heterotic result, one has  to add to the couplings \reef{L2122}  the  fixed couplings at order $\alpha'^2$ which are resulted from the deformed  gauge transformations, \ie $-\frac{\alpha'^2}{12}\Omega_{\mu\nu\alpha}\Omega^{\mu\nu\alpha}$ where $\Omega$ is the three-form Chern-Simons which can be written in terms of spin connection,
\beqa
\Omega_{\mu\nu\alpha}&=&\omega_{[\mu i}{}^j\prt_\nu\omega_{\alpha]j}{}^i+\frac{2}{3}\omega_{[\mu i}{}^j\omega_{\nu j}{}^k\omega_{\alpha]k}{}^i\,\,;\,\,\,\omega_{\mu i}{}^j=\prt_\mu e_\beta{}^j e^\beta{}_i-\Gamma_{\mu\beta}{}^\gamma e_\gamma{}^j e^\beta{}_i
\eeqa
 where $e_\mu{}^ie_\nu{}^j\eta_{ij}=G_{\mu\nu}$. 
 Adding this term, the corresponding T-duality transformations and the 60 parameters in \reef{L2122} may be fixed in the heterotic theory. We didn't perform this calculations, however, one expects all parameters in \reef{L2122} to be zero.   Similar calculation at order $\alpha'$ has been done in \cite{Garousi:2019wgz}. The T-duality structure of the couplings in the heterotic theory  has been studied in \cite{Bedoya:2014pma,Coimbra:2014qaa,Lee:2015kba,Baron:2017dvb,Baron:2018lve} in the DFT formalism. 
 
 There is  another deformation of $B$-field gauge transformations which correspond to the Chiral string theory \cite{Hohm:2013jaa}. The deformation at order $\alpha'$ is the same as the deformation in the heterotic theory in which spin connection is replaced by  the Christoffel connection \cite{Hohm:2014eba}. The low energy effective action of this theory at the leading order,   is given by the T-duality invariant action \reef{S0bf} and  at the order $\alpha'$, it is given by the T-duality invariant coupling   $H^{\mu\nu\alpha}\bf{\Omega}_{\mu\nu\alpha}$ \cite{Hohm:2015doa}\footnote{The parameter $b_1$, in the  parity invariant    action \reef{S1bf}, corresponding to the Chiral string is $b_1=0$.} where the three-form Chern-Simons $\bf{\Omega}_{\mu\nu\alpha}$  is resulted from the deformed gauge transformation, \ie 
\beqa
\bf{\Omega}_{\mu\nu\alpha}&=&\Gamma_{[\mu\beta}{}^\gamma\prt_\nu\Gamma_{\alpha]\gamma}{}^\beta+\frac{2}{3}\Gamma_{[\mu\beta}{}^\gamma\Gamma_{\nu\gamma}{}^\lambda\Gamma_{\alpha]\lambda}{}^\beta
\eeqa
To find the effective action at order $\alpha'^2$, one has  to add to the couplings \reef{L2122}  the  fixed couplings at order $\alpha'^2$, \ie $-\frac{\alpha'^2}{12}\bf{\Omega}_{\mu\nu\alpha}\bf{\Omega}^{\mu\nu\alpha}$.  Adding this term, the corresponding T-duality transformations and the 60 parameters in \reef{L2122} may be fixed in the Chiral string  theory by the T-duality constraint method. It would be interesting to perform this calculation to find the $\alpha'^2$-order terms. This theory has been studied in  the DFT formalism in \cite{Hohm:2015mka,Lescano:2016grn}.

In general, both the diffeomorphisms and the $B$-field gauge transformations  may receive higher derivative deformations in a general gauge invariant higher-derivative theory. One may impose these gauge transformations and the deformed T-duality transformations to study the effective action of a higher-derivative theory which is invariant under the gauge transformations and under  the T-duality transformations.  The effective action at the leading order of $\alpha'$ is given by \reef{S0bf}. At order $\alpha'$, the parity invariant part of the effective action would be more general than the action \reef{S1bf}. It would be interesting to find this effective action. 

 \vskip .3 cm
{\bf Acknowledgments}:   This work is supported by Ferdowsi University of Mashhad under grant  1/50251(1398/04/31).



\begin{thebibliography}{9}

\bibitem{Metsaev:1987zx} 
  R.~R.~Metsaev and A.~A.~Tseytlin,
  Nucl.\ Phys.\ B {\bf 293}, 385 (1987).
  doi:10.1016/0550-3213(87)90077-0
  
\bibitem{Garousi:2019cdn} 
  M.~R.~Garousi and H.~Razaghian,
  arXiv:1905.10800 [hep-th].
\bibitem{Scherk:1974mc} 
  J.~Scherk and J.~H.~Schwarz,
  Phys.\ Lett.\  {\bf 52B}, 347 (1974).
  doi:10.1016/0370-2693(74)90059-8
\bibitem{Yoneya:1974jg} 
  T.~Yoneya,
  Prog.\ Theor.\ Phys.\  {\bf 51}, 1907 (1974).
  doi:10.1143/PTP.51.1907
  
\bibitem{Callan:1985ia} 
  C.~G.~Callan, Jr., E.~J.~Martinec, M.~J.~Perry and D.~Friedan,
  Nucl.\ Phys.\ B {\bf 262}, 593 (1985).
  doi:10.1016/0550-3213(85)90506-1

	\bibitem{Fradkin:1984pq} 
  E.~S.~Fradkin and A.~A.~Tseytlin,
  Phys.\ Lett.\  {\bf 158B}, 316 (1985).
  doi:10.1016/0370-2693(85)91190-6
\bibitem{Fradkin:1985fq} 
  E.~S.~Fradkin and A.~A.~Tseytlin,
  Phys.\ Lett.\  {\bf 160B}, 69 (1985).
  doi:10.1016/0370-2693(85)91468-6


\bibitem{Giveon:1994fu} 
  A.~Giveon, M.~Porrati and E.~Rabinovici,
  Phys.\ Rept.\  {\bf 244}, 77 (1994)
  doi:10.1016/0370-1573(94)90070-1
  [hep-th/9401139].
\bibitem{Alvarez:1994dn} 
  E.~Alvarez, L.~Alvarez-Gaume and Y.~Lozano,
  Nucl.\ Phys.\ Proc.\ Suppl.\  {\bf 41}, 1 (1995)
  doi:10.1016/0920-5632(95)00429-D
  [hep-th/9410237].
	
	
\bibitem{Siegel:1993xq} 
  W.~Siegel,
  Phys.\ Rev.\ D {\bf 47}, 5453 (1993)
  doi:10.1103/PhysRevD.47.5453
  [hep-th/9302036].

\bibitem{Siegel:1993th} 
  W.~Siegel,
  Phys.\ Rev.\ D {\bf 48}, 2826 (1993)
  doi:10.1103/PhysRevD.48.2826
  [hep-th/9305073].
	
\bibitem{Siegel:1993bj} 
  W.~Siegel,
  ``Manifest duality in low-energy superstrings,''
  hep-th/9308133.
	
	
	
\bibitem{Hull:2009mi} 
  C.~Hull and B.~Zwiebach,
  JHEP {\bf 0909}, 099 (2009)
  doi:10.1088/1126-6708/2009/09/099
  [arXiv:0904.4664 [hep-th]].
	 
	
\bibitem{Aldazabal:2013sca} 
  G.~Aldazabal, D.~Marques and C.~Nunez,
  Class.\ Quant.\ Grav.\  {\bf 30}, 163001 (2013)
  doi:10.1088/0264-9381/30/16/163001
  [arXiv:1305.1907 [hep-th]].
	
\bibitem{Hohm:2010pp} 
  O.~Hohm, C.~Hull and B.~Zwiebach,
  JHEP {\bf 1008}, 008 (2010)
  doi:10.1007/JHEP08(2010)008
  [arXiv:1006.4823 [hep-th]].
	
\bibitem{Hohm:2014xsa} 
  O.~Hohm and B.~Zwiebach,
  JHEP {\bf 1411}, 075 (2014)
  doi:10.1007/JHEP11(2014)075
  [arXiv:1407.3803 [hep-th]].
	
\bibitem{Marques:2015vua} 
  D.~Marques and C.~A.~Nunez,
  JHEP {\bf 1510}, 084 (2015)
  doi:10.1007/JHEP10(2015)084
  [arXiv:1507.00652 [hep-th]].

\bibitem{Garousi:2018qes} 
  M.~R.~Garousi,
  Phys.\ Rev.\ D {\bf 98}, no. 6, 066008 (2018)
  doi:10.1103/PhysRevD.98.066008
  [arXiv:1805.08977 [hep-th]].
 
	 

\bibitem{Garousi:2017fbe} 
  M.~R.~Garousi,
  Phys.\ Rept.\  {\bf 702}, 1 (2017)
  doi:10.1016/j.physrep.2017.07.009
  [arXiv:1702.00191 [hep-th]].

\bibitem{Buscher:1987sk} 
  T.~H.~Buscher,
  Phys.\ Lett.\ B {\bf 194}, 59 (1987).
  doi:10.1016/0370-2693(87)90769-6
\bibitem{Buscher:1987qj} 
  T.~H.~Buscher,
  Phys.\ Lett.\ B {\bf 201}, 466 (1988).
  doi:10.1016/0370-2693(88)90602-8
  
  
\bibitem{Tseytlin:1991wr} 
  A.~A.~Tseytlin,
  Mod.\ Phys.\ Lett.\ A {\bf 6}, 1721 (1991).
  doi:10.1142/S021773239100186X
	
\bibitem{Bergshoeff:1995cg} 
  E.~Bergshoeff, B.~Janssen and T.~Ortin,
  Class.\ Quant.\ Grav.\  {\bf 13}, 321 (1996)
  doi:10.1088/0264-9381/13/3/002
  [hep-th/9506156].
	
\bibitem{Kaloper:1997ux} 
  N.~Kaloper and K.~A.~Meissner,
  Phys.\ Rev.\ D {\bf 56}, 7940 (1997)
  doi:10.1103/PhysRevD.56.7940
  [hep-th/9705193].
  
\bibitem{Razaghian:2017okr} 
  H.~Razaghian and M.~R.~Garousi,
  JHEP {\bf 1802}, 056 (2018)
  doi:10.1007/JHEP02(2018)056
  [arXiv:1709.01291 [hep-th]].
	
\bibitem{Razaghian:2018svg} 
  H.~Razaghian and M.~R.~Garousi,
  Phys.\ Rev.\ D {\bf 97}, 106013 (2018)
  doi:10.1103/PhysRevD.97.106013
  [arXiv:1801.06834 [hep-th]].
  
\bibitem{Garousi:2019wgz} 
  M.~R.~Garousi,
  Phys.\ Rev.\ D {\bf 99}, no. 12, 126005 (2019)
  doi:10.1103/PhysRevD.99.126005
  [arXiv:1904.11282 [hep-th]].
	
\bibitem{Bento1990} 
  M.~C.~Bento, O.~Bertolami and J.~C.~Romao,
  Phys.\ Lett.\ B {\bf 252}, 401 (1990).
  doi:10.1016/0370-2693(90)90559-O
	 
	
	 
\bibitem{Meissner:1996sa} 
  K.~A.~Meissner,
  Phys.\ Lett.\ B {\bf 392}, 298 (1997)
  doi:10.1016/S0370-2693(96)01556-0
  [hep-th/9610131].
	
	 
\bibitem{Metsaev:1986yb} 
  R.~R.~Metsaev and A.~A.~Tseytlin,
  Phys.\ Lett.\ B {\bf 185}, 52 (1987).
  doi:10.1016/0370-2693(87)91527-9

	
\bibitem{Nutma:2013zea} 
  T.~Nutma,
  Comput.\ Phys.\ Commun.\  {\bf 185}, 1719 (2014)
  doi:10.1016/j.cpc.2014.02.006
  [arXiv:1308.3493 [cs.SC]].
	
\bibitem{Hohm:2019jgu} 
  O.~Hohm and B.~Zwiebach,
  arXiv:1905.06963 [hep-th].
	
	
  
 
  
\bibitem{Green:1984sg} 
  M.~B.~Green and J.~H.~Schwarz,
  Phys.\ Lett.\  {\bf 149B}, 117 (1984).
  doi:10.1016/0370-2693(84)91565-X
  
\bibitem{Bedoya:2014pma} 
  O.~A.~Bedoya, D.~Marques and C.~Nunez,
  JHEP {\bf 1412}, 074 (2014)
  doi:10.1007/JHEP12(2014)074
  [arXiv:1407.0365 [hep-th]].
\bibitem{Coimbra:2014qaa} 
  A.~Coimbra, R.~Minasian, H.~Triendl and D.~Waldram,
  JHEP {\bf 1411}, 160 (2014)
  doi:10.1007/JHEP11(2014)160
  [arXiv:1407.7542 [hep-th]].
\bibitem{Lee:2015kba} 
  K.~Lee,
  Nucl.\ Phys.\ B {\bf 899}, 594 (2015)
  doi:10.1016/j.nuclphysb.2015.08.013
  [arXiv:1504.00149 [hep-th]].
\bibitem{Baron:2017dvb} 
  W.~H.~Baron, J.~J.~Fernandez-Melgarejo, D.~Marques and C.~Nunez,
  JHEP {\bf 1704}, 078 (2017)
  doi:10.1007/JHEP04(2017)078
  [arXiv:1702.05489 [hep-th]].
\bibitem{Baron:2018lve} 
  W.~H.~Baron, E.~Lescano and D.~Marqués,
  JHEP {\bf 1811}, 160 (2018)
  doi:10.1007/JHEP11(2018)160
  [arXiv:1810.01427 [hep-th]].
  
\bibitem{Hohm:2013jaa} 
  O.~Hohm, W.~Siegel and B.~Zwiebach,
  JHEP {\bf 1402}, 065 (2014)
  doi:10.1007/JHEP02(2014)065
  [arXiv:1306.2970 [hep-th]].
\bibitem{Hohm:2014eba} 
  O.~Hohm and B.~Zwiebach,
  JHEP {\bf 1501}, 012 (2015)
  doi:10.1007/JHEP01(2015)012
  [arXiv:1407.0708 [hep-th]].
\bibitem{Hohm:2015doa} 
  O.~Hohm and B.~Zwiebach,
  JHEP {\bf 1604}, 101 (2016)
  doi:10.1007/JHEP04(2016)101
  [arXiv:1510.00005 [hep-th]].
\bibitem{Hohm:2015mka} 
  O.~Hohm and B.~Zwiebach,
  Phys.\ Rev.\ D {\bf 93}, no. 6, 064035 (2016)
  doi:10.1103/PhysRevD.93.064035
  [arXiv:1509.02930 [hep-th]].
\bibitem{Lescano:2016grn} 
  E.~Lescano and D.~Marques,
  JHEP {\bf 1706}, 104 (2017)
  doi:10.1007/JHEP06(2017)104
  [arXiv:1611.05031 [hep-th]].


 
\end{thebibliography}
\end{document}